\documentstyle[twoside,fleqn,espcrc2]{article}

\catcode`@=11
\def\@sim#1#2{\setbox0=\hbox{$\sim$}\lower.9\ht0\vbox{\baselineskip0pt
              \lineskip0.1ex\ialign{$\m@th#1\hfill##\hfill$\crcr#2\crcr
              \sim\crcr}}}
\def\lsim{\mathrel{\mathpalette\@sim<}}

\input epsf

\title{Scaling topological charge in the ${\rm CP}^{3}$ model using a
fixed point action} 

\author{Rudolf Burkhalter \address{Institute for Theoretical Physics,%
        University of Berne, \\ 
        Sidlerstrasse 5, CH-3012 Berne, Switzerland}}

\begin{document}

\begin{abstract}
We define a fixed point action in two-dimensional lattice ${\rm CP}^{N-1}$
models.  The fixed point action is a classical perfect lattice action, which
is expected to show strongly reduced cut-off effects in numerical simulations.
Furthermore, the action has scale invariant instanton solutions, which enables
us to define a topological charge without topological defects.  We present
results for the scaling of the topological suceptibility from a Monte Carlo
simulation in the ${\rm CP}^{3}$ model.
\end{abstract}

\maketitle

\section{Introduction}

The use of fixed point (FP) actions for asymptotically free theories has been
shown to be profitable in reducing lattice artefacts in numerical
simulations~\cite{HASENFRATZ,DEGRAND}. Furthermore, it has been shown how to
define a FP topological charge in order to study topological
properties~\cite{BLATTER}. However, a determination of the topological
susceptibility in the O(3) nonlinear $\sigma$-model did not show a
scaling behavior. The result suggests, not quite unexpectedly~\cite{SCHWAB},
that the topological susceptibility is not a physical quantity in this
model. In the ${\rm CP}^{3}$ model one expects the topological susceptibility
to be meaningful. However, recent determinations gave contradictory
results~\cite{CAMPOSTRINI,WOLFF,HASENBUSCH,JANSEN,IRVING}.

In order to overcome these problems, we discuss the definition of a fixed
point action and a fixed point topological charge for ${\rm CP}^{N-1}$ models.
We present results of numerical simulations in the ${\rm CP}^{3}$ model. A
more detailed discussion of the definitions and the numerical results can be
found in Ref.~\cite{BURKHALTER}.

\section{Fixed point action}

Two-dimensional ${\rm CP}^{N-1}$ models are a class of models which in many
respects are similar to QCD. They consist of $N$-component, complex spin
fields $z^i(x)$ of unit length. The FP action is defined as the
FP of an exact renormalization group (RG) transformation. The RG
transformation relates the spins $z_n$ sitting at the lattice sites of a
$2\times 2$ block $n_B$ of a fine lattice with a block spin $\zeta_{n_B}$
sitting on a coarse lattice. The FP action is determined by the FP equation
\begin{equation}
  {\cal A}_{\mbox{\tiny FP}}(\zeta) = \min_{\{z\}}\left\{{\cal A}_{\mbox{\tiny
  FP}}(z) + {\cal T}(\zeta,z) \right\}, 
  \label{FP-equation}
\end{equation}
with the transformation kernel 
\begin{equation}
  {\cal T}(\zeta,z) = \kappa \sum_{n_{B}} \left (  \hat{\lambda}_{n_B} - 
\sum_{n \in n_B} |\zeta_{n_B} \bar{z}_{n}|^2 \right ).  
  \label{kernel}
\end{equation}
Here $\kappa$ is a free parameter that is used to optimize the corresponding
FP action and $\hat{\lambda}_{n_B}$ is the largest eigenvalue of the matrix
\begin{equation} 
  M_{n_B} = \sum_{n \in n_B} z_n \otimes \bar{z}_n.
  \label{matrix}
\end{equation}
One can solve the FP equation iteratively for any given input configuration
$\{\zeta\}$. In doing this, we may use a multigrid with $\{\zeta\}$ on the
coarsest level. The configurations on the finer levels are varied until the
minimum is reached which yields the value of the the FP action for
$\{\zeta\}$. On the finest level the field is very smooth, so we can use any
discretization of the continuum action, e.g. the standard action.

If one wants to use the FP action in a numerical simulation one has to use a
parametrization. Our parametrization has the form
\begin{eqnarray}
  {\cal A}_{\mbox{\tiny FP}}^{\rm par}(z)\!\!\!\! 
  &=& \!\!\! -{1\over 2} \sum_{n,r} \rho(r)
      \theta_{n,n+r}^2 \label{parametrization} \\
  &+& \!\!\!\!\! \sum_{n_i,n_j,\ldots} \!\!\!\! \mbox{ coupling }\!\times\! 
      \mbox{ products of } \theta_{n_i,n_j}^2, \nonumber 
\end{eqnarray}
where $\theta_{n_i,n_j} = \arccos\left(\left| \bar{z}_{n_i}
z_{n_j}\right|\right)$ is the angle between two spins. We determined a
parametrization for the ${\rm CP}^{3}$ model with 32 couplings (2 analytically
calculated quadratic couplings $\rho$ and 30 numerically determined higher
order couplings which involve 2-, 3- and 4-spin interactions) which deviates
only negligibly from the (minimized) FP action even for rough configurations
\cite{BURKHALTER}.

\section{Fixed point topological charge}

In Ref.~\cite{BLATTER} it was shown, that the FP action admits scale invariant
instanton solutions. In particular, it can be shown, that if $\{\zeta\}$ is an
instanton configuration of size $\hat\rho$ (in units of the lattice spacing)
and hence its action has a value of $2\pi$, then the minimizing configuration
$\{z(\zeta)\}$ is an instanton configuration as well. Its size is $2\hat\rho$
and the value of its action is also $2\pi$.

This property leads us to the definition of a FP topological
charge~\cite{BURKHALTER}. It is obtained in the limit of infinitely many RG
transformations:
\begin{equation}
  Q_{\mbox{\rm \tiny FP}}(\zeta) = \lim_{k \rightarrow \infty}
  Q(z^{(k)}(\zeta)),
\end{equation}
where $\{ z^{(k)}\}$ is the solution of the iterated FP~equation on the lowest
level in a $k$-level multigrid.  As the configurations get smoother on each
successive level, one may choose for the charge $Q$ on the lowest level any
lattice discretization of the topological charge. In this work we used the
geometric definition of the topological charge.  The combination FP action and
FP topological charge obeys for any configuration the inequality ${\cal
A}_{\mbox{\rm \tiny FP}} \geq 2 \pi \, | Q_{\mbox{\rm \tiny FP}} |$,
i.e. there are no topological defects.

\begin{figure}[htb]
\vspace{9pt}
\epsfxsize=62mm
\epsfysize=48mm
\hspace{6mm}
\epsfbox{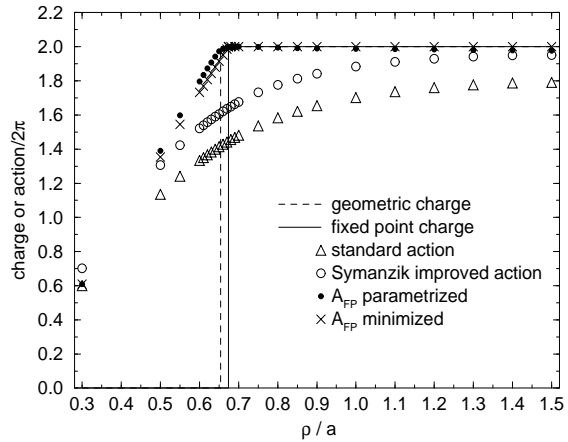}
\caption{Actions and charge of instantons with radii of the order of one
lattice spacing.}
\label{fig:inst}
\end{figure}

This property is well illustrated in Fig.~\ref{fig:inst}. Here we plot for
the ${\rm CP}^{3}$ model the topological charges and the actions for
2-instanton solutions on a torus.

To calculate the FP charge in numerical simulations, it is very time consuming
to minimize a multigrid for 
every configuration.  One needs a
parametrization of the solution $\{z(\zeta)\}$ of the FP~equation. We
construct a gauge invariant parametrization by calculating the matrix
\begin{eqnarray}
  M_n\!\!\!\! &=& \!\!\! \sum_{n_B^{}} \alpha(n,n_B^{})\;\zeta_{n_B^{}}\!\!
                  \otimes\bar\zeta_{n_B^{}} \label{parametfield}\\
      &+& \!\!\!\!\!\!\!\! \sum_{n_B^{}
       \atop m_B^{},m_{B}'} \!\!\!\! \beta(n,n_{\mbox{\tiny
       B}}^{},m_{\mbox{\tiny B}}^{},m_{\mbox{\tiny B}}') 
       \;\theta^2_{m_B^{},m_B'} \;\zeta_{n_B^{}}\!\!\otimes \bar\zeta_{n_B^{}},
       \nonumber
\end{eqnarray}
and define the fine field variable $z_n$ as the eigenvector of $M_n$ with
largest eigenvalue. In higher order terms enters the angle
$\theta^2_{m_B^{},m_B'}$ between the coarse spins at sites $m_B$ and $m_B'$,
respectively. Similar to the FP action we can determine for the ${\rm CP}^{3}$
model the coefficients $\alpha$ and $\beta$ analytically and by a fitting
procedure~\cite{BURKHALTER}.

\section{Numerical results}

We performed MC simulations in the ${\rm CP}^{3}$ model using the parametrized
FP action with a hybrid overrelaxation algorithm. We determined the FP charge
by calculating the charge on the first finer level which was obtained by
either minimizing Eq.~(\ref{FP-equation}) or by using the
parametrization~(\ref{parametfield}). We keep the volume $L/\xi \simeq 5.5 -
6$ approximately constant. At this value we still observe finite size effects
but the constant volume allows to look for a scaling behavior.

\begin{figure}[htb]
\vspace{9pt}
\epsfxsize=62mm
\epsfysize=48mm
\hspace{6mm}
\epsfbox{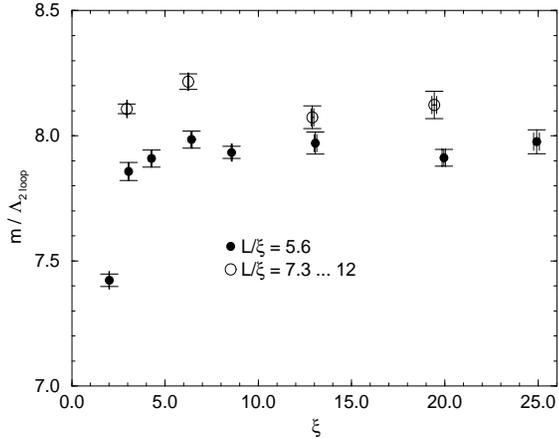}
\caption{Asymptotic scaling test for the mass in the ${\rm CP}^{3}$ model.}
\label{fig:asympt}
\end{figure}

The mass gap shows an unexpected asymptotic scaling behavior even at quite
small correlation lengths (Fig.~\ref{fig:asympt}). Furthermore the ratio
$m/\Lambda^{(2)}_L = 8.1(1)$ (obtained in ``infinite'' volume) is remarkably
small.

\begin{figure}[htb]
\vspace{9pt}
\epsfxsize=62mm
\epsfysize=48mm
\hspace{6mm}
\epsfbox{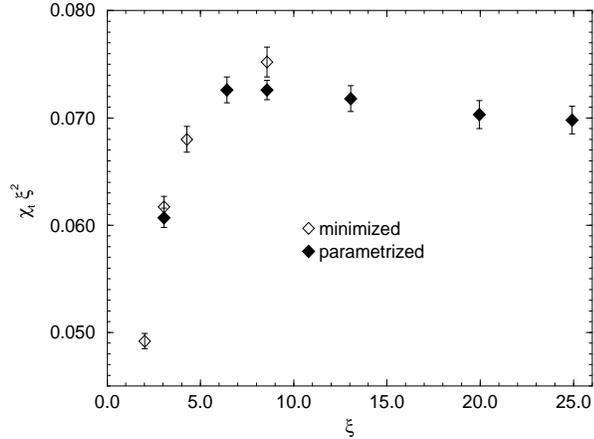}
\caption{Scaling test for the topological susceptibility. These are the
results measured in volumes $L / \xi \simeq 5.6$.}
\label{fig:suscres}
\end{figure}

For the topological susceptibility
\begin{equation}
  \chi_t = \frac{\langle Q^2\rangle}{V},
\end{equation}
we observe a scaling behavior. In Fig.~\ref{fig:suscres} we show the results
for the dimensionless quantity $\chi_t\, \xi^2$. One sees a raise at small
correlation lengths, which is due to lost small instantons in this region. At
correlation length $\xi \simeq 10$ this effect is already saturated and a
scaling plateau is reached. In ``infinite'' volume we obtain the value $\chi_t
\cdot \xi^2 = 0.070(2)$.

\end{document}